\UseRawInputEncoding
\pdfoutput=1
\documentclass{iopjournal}

\begin{document}

\articletype{Paper} 

\title{Development of Low-Noise Two-stage dc-SQUID for TES Detector Readout}

\author{
Nan Li$^1$
\orcid{0000-0003-4915-349X}, 
Mengjie Song$^1$
\orcid{0009-0004-7695-4799}, 
Sixiao Hu$^1$
\orcid{0009-0006-5305-2981}, 
Wentao Wu$^2$
\orcid{0000-0002-6769-0267}, 
Songqing Liu$^{3,4}$
\orcid{0000-0000-0000-0000}, 
Tangchong Kuang$^5$
\orcid{0009-0003-3904-7551}, 
Yudong Gu$^1$
\orcid{0000-0002-4918-7553}, 
Xiangxiang Ren$^5$
\orcid{0000-0001-8678-2696}, 
Xufang Li$^1$
\orcid{0000-0002-2793-9857}, 
He Gao$^{1,*}$
\orcid{0009-0002-6013-1953}, 
Zhengwei Li$^{1,*}$
\orcid{0000-0002-4888-0858}, 
and 
Congzhan Liu$^{1,*}$
\orcid{0000-0002-4834-9637}}

\affil{$^1$State Key Laboratory of Particle Astrophysics, Institute of High
Energy Physics, CAS, 19B Yuquan Road, Shijingshan District,
100049, Beijing, China}

\affil{$^2$State Key Laboratory of Materials for Integrated Circuits, Shanghai Institute of Microsystem and Information Technology, Chinese Academy of Sciences, Shanghai 200050, China}

\affil{$^3$Shandong University, Jinan, Shandong 250100, China}

\affil{$^4$Shandong Institute of Advanced Technology (SDIAT), Jinan, Shandong 250100, China}

\affil{$^5$Key Laboratory of Particle Physics and Particle Irradiation (MOE), Institute of Frontier and Interdisciplinary Science, Shandong University, Qingdao, Shandong, 266237, China}

\affil{$^*$Author to whom any correspondence should be addressed.}

\email{
hgao@ihep.ac.cn,
lizw@ihep.ac.cn
and
liucz@ihep.ac.cn
}


\begin{abstract}
Direct-current superconducting quantum interference devices (dc-SQUIDs) are one of the most sensitive magnetic detectors.
These sensors are extensively used in the readout of superconducting transition edge sensors (TESs), which are used for the detection of weak signals.
A cosmic microwave background (CMB) polarization telescope operating in 22-48 GHz is currently under developing. 
The TESs calorimeter of the telescope will be readout by a time-division multiplexer (TDM) SQUID readout system.
We develop a two-stage dc-SQUID amplifier circuit, comprising an input-stage SQUID with 4 SQUID cells and a series SQUID array (SSA) with 100 SQUID cells. 
This configuration has been shown to achieve extremely high signal gain while effectively controlling system noise.
We assess the system noise at $300$ $mK$ in an adiabatic demagnetisation refrigerator (ADR).
The the measured magnetic flux noise of the two-stage SQUID circuit system is approximately $0.3$ ${\mu}\Phi_{0}/\sqrt{Hz}$ at $10$ $kHz$.
The current noise equivalent to the input coil of input SQUID is about $2.4$ $pA/\sqrt{Hz}$. 
This result meets the low-noise readout requirements of the CMB TES and other applications with TES detectors.
\end{abstract}

\section{Introduction}
Transition-edge sensors (TESs) have been applied in numerous fields due to extremely low thermal noise\cite{li1998,1995NEWGR,2409.05643,Liu2024}.
TESs are widely used in weak signal detection experiments, such as millimeter wave\cite{li2017tibet}, X-ray\cite{gottardi2021review,shuo2021development}, $\gamma$-ray\cite{zhang2022transition}, cold dark matter\cite{cang2020probing}, neutrinoless double beta decay\cite{bratrud2024first}, and so on.
The Ali primordial gravitational wave detection project is developing TES detection techonlogy to search primordial gravitional waves\cite{weiland2022polarized}. 
Due to the influence of Galactic foreground, B-mode polarization from primordial gravitational waves has not yet been accurately observed~\cite{adam2016planck}. 
The Ali primordial gravitational wave detection project is designing a telescope to detect CMB polarization in the frequency range of 22 to 48 GHz (AliCPT-40G) to study Galactic foreground.
The AliCPT-40G telescope will deploy approximately 600 TES detectors on the focal plane, utilizing a time-division multiplexing (TDM) electronic system based on two-stage direct-current superconducting quantum interference devices (dc-SQUIDs).
Dc-SQUID has ultra sensitive responses to weak magnetic signals and notably low noise performance, which is suitable for TES
readout\cite{irwin2021squids,feng2025development,kiviranta2023components}.
The working state of TES is near superconducting\cite{irwin2005transition}, which enables sensitive response to weak signals.
TES exhibits extremely low thermal noise, typically ranging from tens to hundreds of $pA/\sqrt{Hz}$ with excess noise\cite{ullom2004suppression,luukanen2003fluctuation,ullom2004characterization,wessels2021model}.
Thus in order to ensure the high performance of TES, it is necessary to control the total noise of the two-stage dc-SQUID readout system below $10$ $pA/\sqrt{Hz}$, thereby preventing it from becoming the major system noise source.

We have developed a two-stage dc-SQUID circuit capable of reading out CMB TES calorimeters and compatible with other most TES applications.
The paper is organized as follows:
Section~\ref{sec:SQ_design} describes the chips design of the input SQUID and SSA.
Section~\ref{sec:SQ_chara} presents the measurements of design parameters and system performance.
Section~\ref{sec:conclusion} provides the conclusions of the paper.

\section{Input SQUID and SSA design}
\label{sec:SQ_design}
\begin{figure}[htbp]
    \centering
    \includegraphics[width=.8\textwidth]{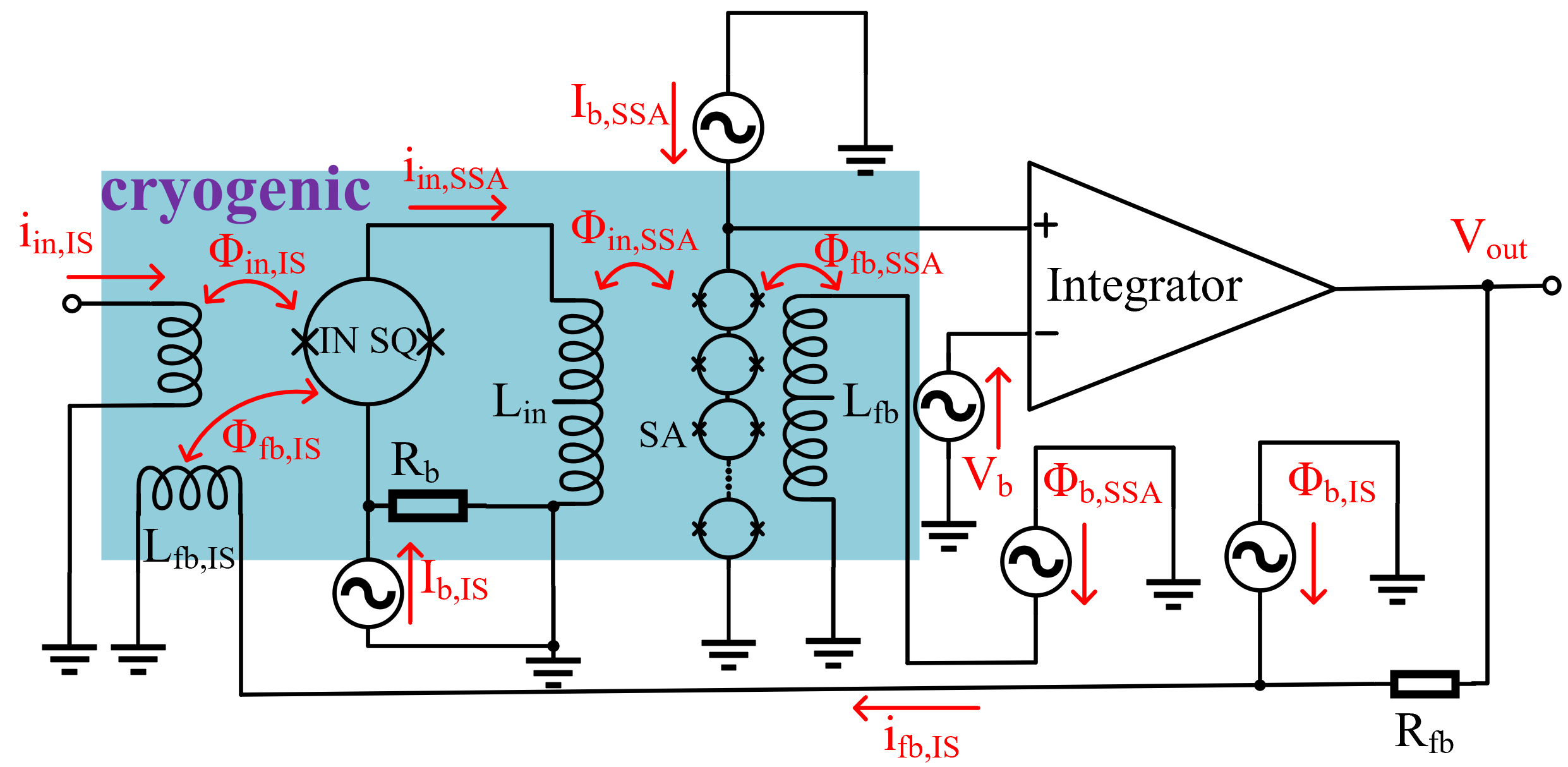}
    \caption{Two-stage dc-SQUID circuit for TES readout.
    The current signal $i_{in,IS}$ flowing through the TES is converted into a magnetic flux signal $\Phi_{in.IS}$ by the input coil $L_{in.IS}$ of the input SQUID. This flux signal cancels the feedback magnetic flux signal $\Phi_{fb.IS}$. Flux-locked loop linearly amplifies the TES signal based on the room-temperature integrating amplifier.
    \label{fig:fig0_circuit}}
\end{figure}

A two-stage dc-SQUID, comprising an input SQUID formed by one or several SQUID cells in series and a series SQUID array (SSA) with numerous cascaded SQUID cells, combined with an integrating amplifier circuit, forms a flux-locked loop circuit \cite{welty2002two,cantor1997low, Zhang2025fabrication} can achieve extremely high signal gain while effectively controlling system noise, as shown in Figure~\ref{fig:fig0_circuit}.
The blue area in Figure~\ref{fig:fig0_circuit} indicates the low-temperature region, while the red arrow denotes the primary signal transmission paths and the locking point tuning signals.
The input SQUID loop exhibits a high current sensitivity relative to the input coil, which allows for more effective amplification of the weak current signals flowing through the TES\cite{matthias2025highly}.
The primary noise sources in the readout system are input SQUID and SSA noise and room-temperature amplifier noise.
The large flux conversion coefficient $I_{\Phi}$ effectively reduces noise contributions from the SSA to the TES.
After further amplification by the SSA, the signal amplitude can reach hundreds of ${\mu}V$ or even several $mV$, ultimately amplified again by the room-temperature amplifier for final readout.
SSA with numerous SQUID cells, can exhibit a very large flux conversion coefficient $V_{\Phi}$, which effectively reduces the noise contribution of the room-temperature amplifier to the system\cite{wu2022development}.

\begin{figure}[htbp]
    \centering
    \includegraphics[width=.8\textwidth]{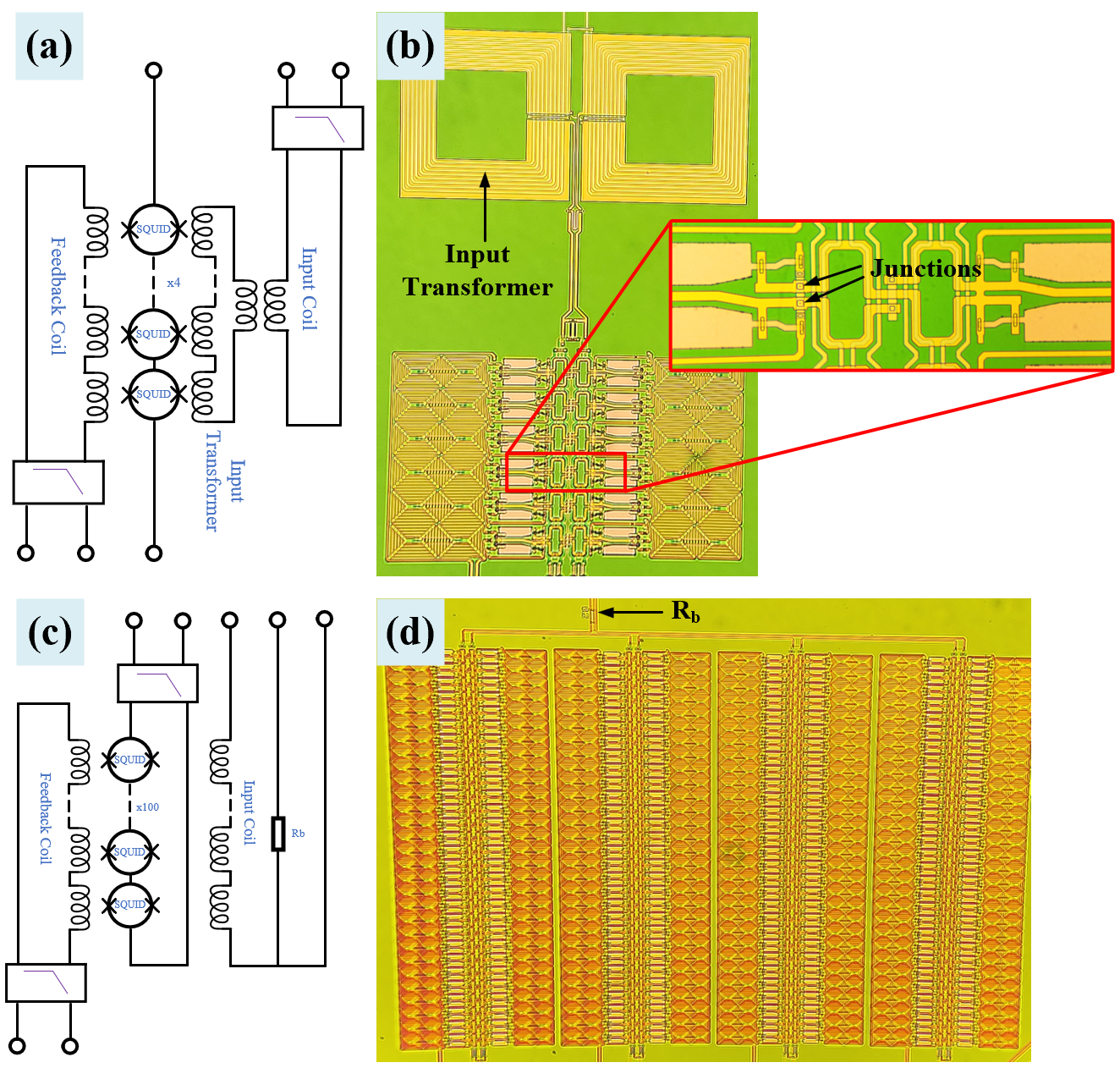}
    \caption{Schematic of the dc-SQUID chips. (a) Input SQUID schematic; (b) Input SQUID board layout; (c) SSA schematic; (d) SSA board layout.
    \label{fig:fig1_IS_des}}
\end{figure}

The input SQUID is designed as double-transformer type~\cite{muhlfelder2003double,polushkin2002tightly,drung2007highly}, as shown in Figure~\ref{fig:fig1_IS_des} (a),(b), it consists of 4 real SQUID cells, flanked by 2 dummy SQUID cells at either end. 
Each cell is a first-order series gradiometer formed by two nearly rectangular washers. 
The hole dimension of each washer is 40 ${\mu}m$ $\times$ 12 ${\mu}m$. 
The input coil, running on top of both washers, is coupled to an input transformer so that the relatively low SQUID inductance (140 $pH$) can be matched with a relatively high input inductance. 
The feedback coil is located at both sides of the washers. 
On-chip low-pass filters are added to the coils. 
The Josephson junction size of input SQUID is 2.5 ${\mu}m$ $\times$ 2.5 ${\mu}m$, cooling fins~\cite{falferi2008cooling} are connected to the shunt resistors of junctions to enlarge the cooling area for minimizing the hot electron effect below 1 $K$. 
The asymmetric bias injection technique~\cite{uehara1993asymmetric} were used in each SQUID cell, and a lot of dummy structures were added to enhance the gradiometry of SQUID:
The SSA consists of a 100 SQUID cells (4 rows $\times$ 25 real SQUID cells), as shown in Figure~\ref{fig:fig1_IS_des} (c),(d).
Each cell is designed as a first-order series gradiometer as well. 
It is highly similar to the input SQUID cell, except for the junction size of 3 ${\mu}m$ $\times$ 3 ${\mu}m$ and each washer hole dimensions of 40 ${\mu}m$ $\times$ 9 ${\mu}m$, and there are also low-pass filters near the output pads. 
The detailed design parameters of input SQUID and SSA are summarized in Table~\ref{tab:IS_para} and~\ref{tab:SSA_para}. 
The proposed input SQUID and SSA devices were fabricated with high-quality Nb/Al–AlOx/Nb trilayer JJs in a standard fabrication process at the Superconducting Electronics Facility (SELF), of the Shanghai Institute of Microsystem and Information Technology (SIMIT). The Nb/Al-AlOx/Nb trilayer is deposited in situ using magnetron sputtering with a critical current density $J_c$ of about 1.3 ${\mu}A/{\mu}m^2$ (at 300 $mK$) and a specific capacity Cs of 40.5 $fF/{\mu}m^2$. The detailed process steps have been reported in previous works~\cite{wu2022development,ying2021development}.
\begin{table}[htbp]
    \centering
    \caption{Design parameters of the input SQUID at 300 $mK$.\label{tab:IS_para}}
    \begin{tabular}{|c|c|}
    \hline
    Parameters & Designed value\\
    \hline
    Critical current of input SQUID $2I_{c0,IS}$ & $16$ ${\mu}A$\\
    \hline
    SQUID loop inductance $L_{s,IS}$ & $140$ $pH$\\
    \hline
    $\beta_c$ of Junctions& $0.23$\\
    \hline
    $\beta_L$ of the input SQUID cell  & $1.10$\\
    \hline
    Input coil inductance $L_{in,IS}$ & $39.6$ $nH$\\
    \hline
    Current sensitivity of Input SQUID $1/{M_{in,IS}}$ & $8.1$ ${\mu}A/\Phi_0$\\
    \hline
    Feedback coil coupling $1/{M_{fb,IS}}$ & $39.7$ ${\mu}A/\Phi_0$\\
    \hline
    \end{tabular}
\end{table} 
\begin{table}[htbp]
    \centering
    \caption{Design parameters of SSA at 300 $mK$.\label{tab:SSA_para}}
    \begin{tabular}{|c|c|}
    \hline
    Parameters & Designed value\\
    \hline
    Critical current of SSA $2I_{c0,SSA}$ & $23$ ${\mu}A$\\
    \hline
    SQUID loop inductance $L_{s,SSA}$ & $120$ $pH$\\
    \hline
    $\beta_c$ of Junctions& $0.26$\\
    \hline
    $\beta_L$ of the SAA cell  & $1.36$\\
    \hline
    Current sensitivity of SSA $1/{M_{in,SSA}}$ & $27.4$ ${\mu}A/\Phi_0$\\
    \hline
    Feedback coil coupling $1/{M_{fb,IS}}$ & $37.6$ ${\mu}A/\Phi_0$\\
    \hline
    \end{tabular}
\end{table}

\section{Measurement and discussion}
\label{sec:SQ_chara}
System noise is the most significant factor affecting the performance of TES.
The system noise contributions in a two-stage dc-SQUID primarily include intrinsic flux noise from the input SQUID and SSA, as well as noise from room-temperature readout electronics.
Based on the circuit signal transmission model of the two-stage dc-SQUID (Figure~\ref{fig:fig0_circuit}), the system noise $NEI_{sys}$ can be characterized by the equivalent current noise density of the input coil connected to the input SQUID, as shown in Equation~\ref{eq0}.
\begin{equation}
\label{eq0}
    NEI_{sys} = \sqrt{NEI_{IS}^2 + NEI_{SSA}^2 + NEI_{elec}^2}.
\end{equation}
The current sensitivity between the the input coil of input SQUID and its junction loop determines the gain for converting the current signal flowing through the TES into the magnetic flux signal. Therefore, it is a crucial parameter affecting the transmission of system noise.
The transfer function relating the intrinsic flux noise density $n_{IS}$ of the input SQUID to the equivalent current noise density $NEI_{IS}$ refering to its input coil is given by Equation~\ref{eq1}.
\begin{equation}
\label{eq1}
    NEI_{IS} = \frac{n_{IS}}{M_{in,IS}}.
\end{equation}
Additionally, the flux conversion coefficients $I_\Phi$ of input SQUID and $V_\Phi$ of SSA are equally important parameters affecting the system noise. 
The contribution of SSA to the system noise $NEI_{SSA}$ can be converted based on its intrinsic flux noise $n_{SSA}$, as shown in Equation~\ref{eq2}.
\begin{equation}
\label{eq2}
    NEI_{SSA} = \frac{n_{SSA}}{M_{in,SSA}I_{\Phi}M_{in,IS}}.
\end{equation}
The $NEI_{SSA}$ is inversely correlated with the flux conversion coefficient $I_{\Phi}$ which is strongly dependent on the working point~\cite{wu2022development}.
It is evident that an increase in the flux conversion coefficient results in a reduction in the contribution of back-end noise.
Room-temperature electronic noise undergoes further attenuation via the SSA's magnetic flux conversion coefficient $V_{\Phi}$, resulting in a lower system contribution $NEI_{elec}$, as shown in Equation~\ref{eq3}.
\begin{equation}
\label{eq3}
    NEI_{elec} = \frac{n_{elec}}{V_{\Phi}M_{in,SSA}I_{\Phi}M_{in,IS}}.
\end{equation}
Room-temperature electronic noise $n_{elec}$ primarily originates from the voltage noise and current noise of the preamplifier.

\begin{figure}[htbp]
    \centering
    \includegraphics[width=.7\textwidth]{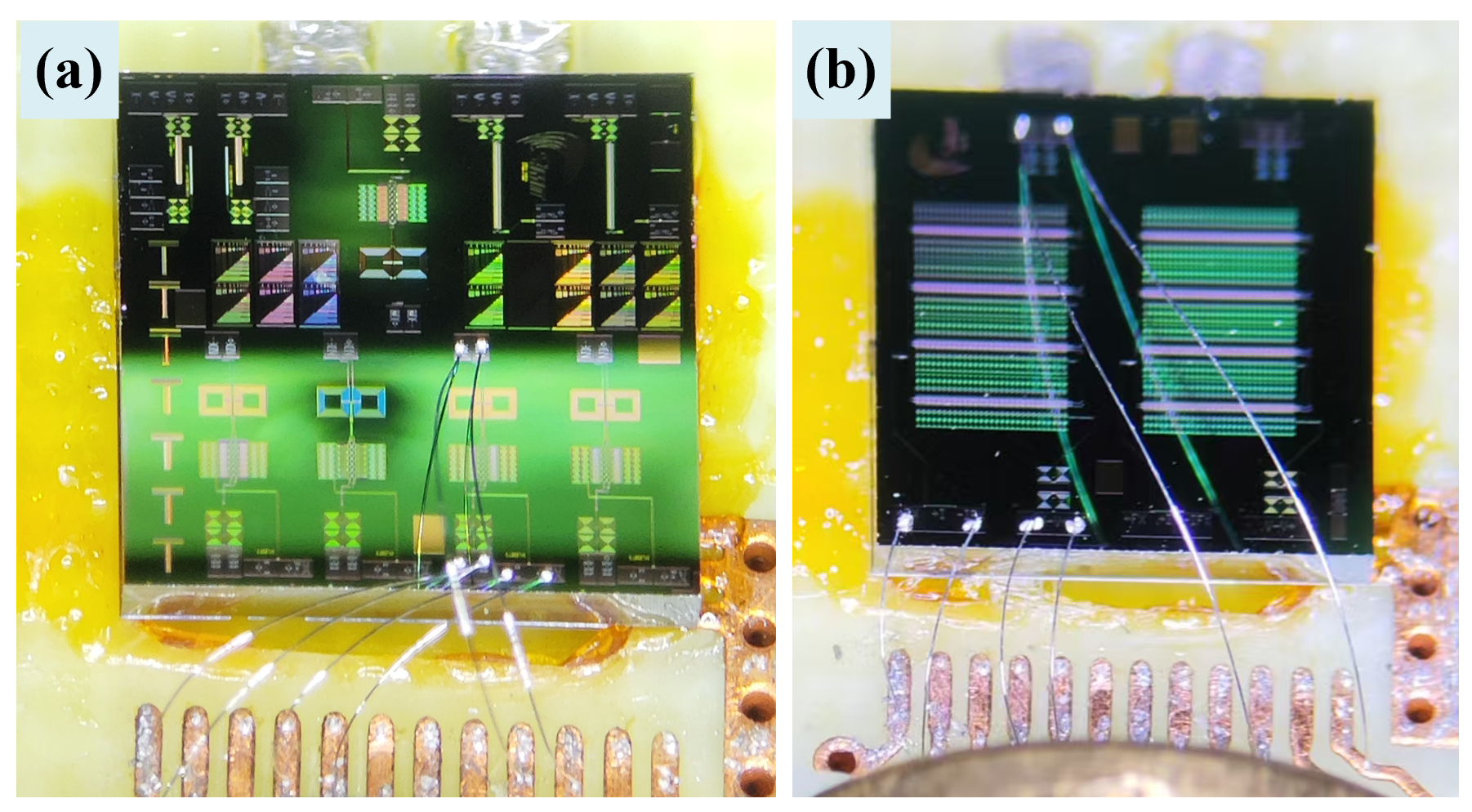}
    \caption{Input SQUID (a) and SSA (b) are connected to pads of PCB via wire-bonding. They are installed on the $300$ $mK$ copper plate in ADR.
    \label{fig:fig2_bond}}
\end{figure}

The current sensitivities, $V_{\Phi}$ and $I_\Phi$ can be obtained by measuring the magnetic flux response $V$-$\Phi$ curves of the SQUIDs.
The $V$-$\Phi$ characteristics of the fabricated input SQUID and SSA chips were measured at $300$ $mK$ within ADR system.
The input SQUID and SSA were each glued to a printed circuit board (PCB) and mounted on the $300$ $mK$ gold-plated copper cold plate in ADR.
Each pad on the chips was connected to the PCB by aluminum bonding wires, as illustrated in Figure~\ref{fig:fig2_bond}.
The chips were finally equipped with a
superconducting Nb can as a shield from stray magnetic fields.

\begin{figure}[htbp]
    \centering
    \includegraphics[width=.8\textwidth]{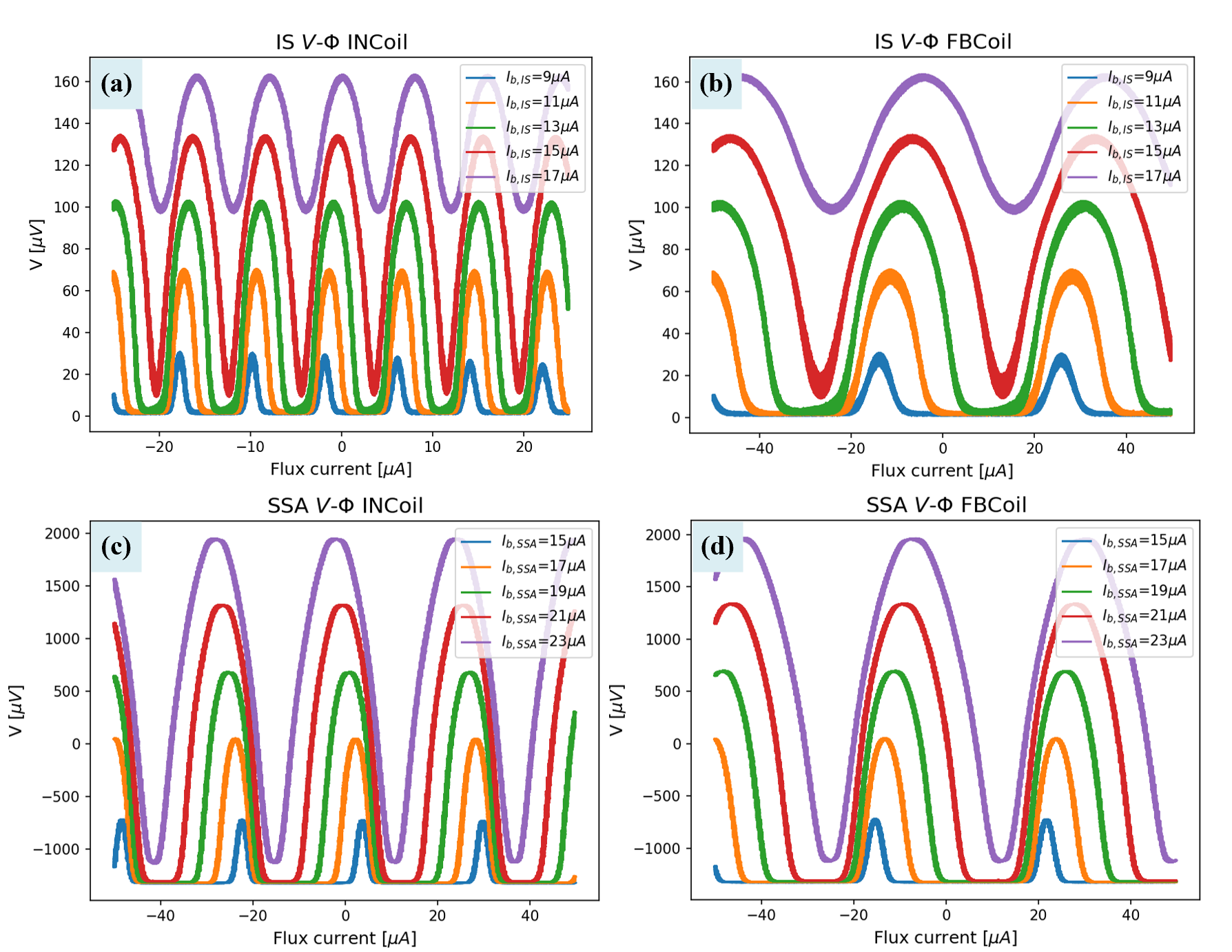}
    \caption{The $V$–$\Phi$ characteristics of the input SQUID and SSA were measured at $300$ $mK$ in ADR. The results correspond to the input coil (a) and feedback coil (b) of the input SQUID, and the input coil (c) and feedback coil (d) of the SSA.
    \label{fig:fig3_Vphi}}
\end{figure}

Magnicon electronics provide all bias currents and coil reference flux inputs to input SQUID and SSA, respectively.
The $V$-$\Phi$ response of the input SQUID and SSA measured under different bias currents is shown in Figure~\ref{fig:fig3_Vphi}.
It can be seen that the current sensitivity between the junction loop and the input coil, as well as the feedback coil, of input SQUID are $8$ ${\mu}A/\Phi_0$ and $40$ ${\mu}A/\Phi_0$ respectively.
The current sensitivity between the SSA's loop and the input coil and feedback coil are $27$ ${\mu}A/\Phi_0$ and $38$ ${\mu}A/\Phi_0$ respectively.
The input SQUID and SSA exhibit maximum voltage swings and maximum flux conversion coefficients near $15$ ${\mu}A$ and $23$ ${\mu}A$ current bias points respectively, which is proximate to the expected critical currents ($2I_{c0,IS}$ and $2I_{c0,SSA}$).

In order to achieve enhanced system noise performance, the two-stage dc-SQUID was bonded (as shown in Figure 4) and also placed at a temperature of $300$ $mK$ in ADR. 
The measured results were based on ultra low-noise Magnicon flux-locked loop readout circuit, which are of $0.33$ $nV/\sqrt{Hz}$ voltage noise and $2.6$ $pA/\sqrt{Hz}$ current noise\cite{drung2006low}.
\begin{figure}[htbp]
    \centering
    \includegraphics[width=.3\textwidth]{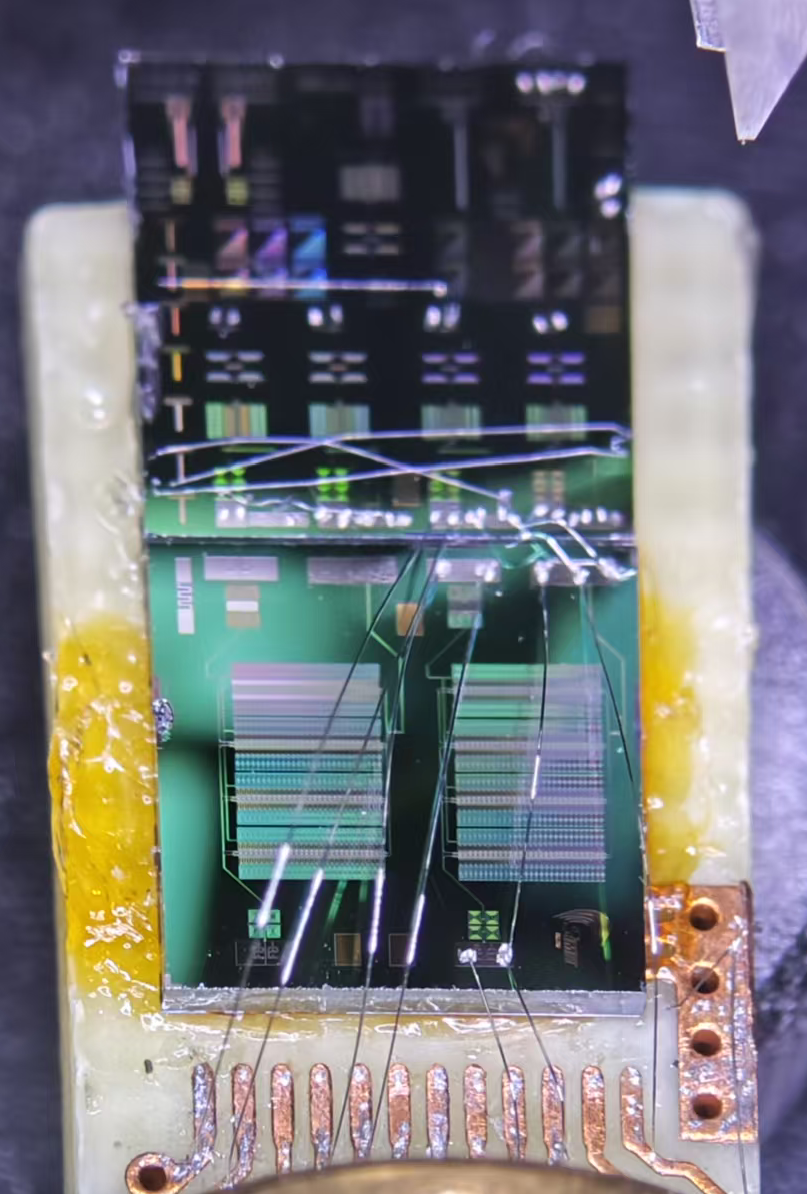}
    \caption{The input SQUID is connected to the SSA via wire-bonding. They are installed on the Nb-shielded PCB of Magnicon at the $300$ $mK$ copper plate in ADR.
    \label{fig:fig4_bond}}
\end{figure}
The SSA exhibits a maximum voltage swing at $300$ $mK$, which is approximately $3$ $mV$ at a current bias of $23$ ${\mu}A$. 
At the optimal working point, the flux conversion coefficient $V_{\phi}$ has the maximum value, which is approximately $10$ $mV/\Phi_0$, as shown in Figure~\ref{fig:fig5_lock_point} (a).
In a two-stage dc-SQUID circuit, the dynamic resistance of the input SQUID is substantially greater than the parallel resistor $R_b$, which is $0.2$ $\Omega$ in our design. 
According to Thevenin's theorem, the bias current will be equivalent to a voltage bias (as shown in Figure~\ref{fig:fig0_circuit}). 
Under a $4$ ${\mu}V$ voltage bias, the maximum current swing measured through the input SQUID is approximately $7$ ${\mu}A$, which is primarily constrained by the magnitude of the approximately linear gain interval ${\Delta}I_{in,SSA}$ around the SSA working point. 
The maximum value of $I_{\Phi}$, at the working (locking) point, is approximately $45$ ${\mu}A/\Phi_0$, as shown in Figure~\ref{fig:fig5_lock_point} (b).
\begin{figure}[htbp]
    \centering
    \includegraphics[width=.7\textwidth]{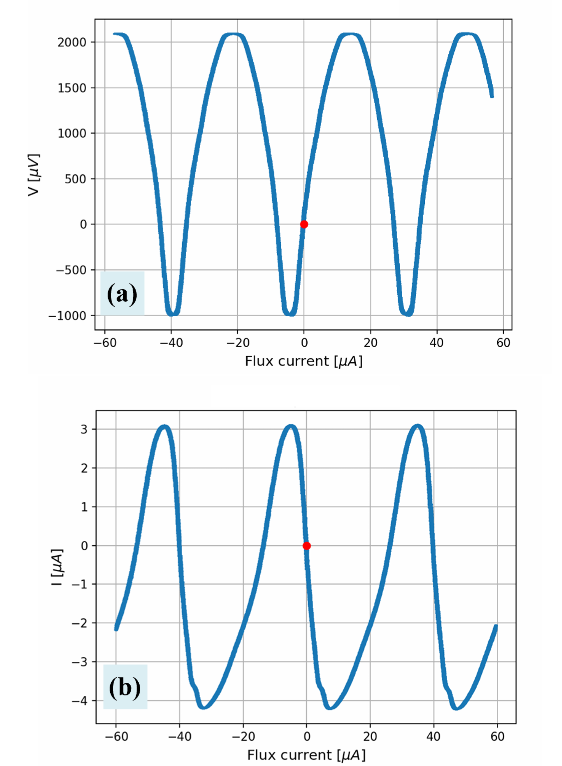}
    \caption{Tuning of the woking points for the SSA (a) and input SQUID (b) with feedback coils. The maximum voltage swing of the SSA is approximately $3$ $mV$ and the maximum $V_\Phi$ is about $10$ $mV/\Phi_0$. The maximum current swing of the input SQUID is approximately $7$ ${\mu}A$ and the maximum $I_\Phi$ is about $45$ ${\mu}A/\Phi_0$ at $300$ $mK$. The red dots indicate the working points.
    \label{fig:fig5_lock_point}}
\end{figure}
The input SQUID $V$-$\Phi$ curve has minor distortion at its base. 
One possible explanation is that the $\beta_c$ exceeds the designed value, thereby inducing Josephson high-frequency oscillation with microstrip resonators\cite{kawakami2002josephson}.

\begin{figure}[htbp]
    \centering
    \includegraphics[width=.7\textwidth]{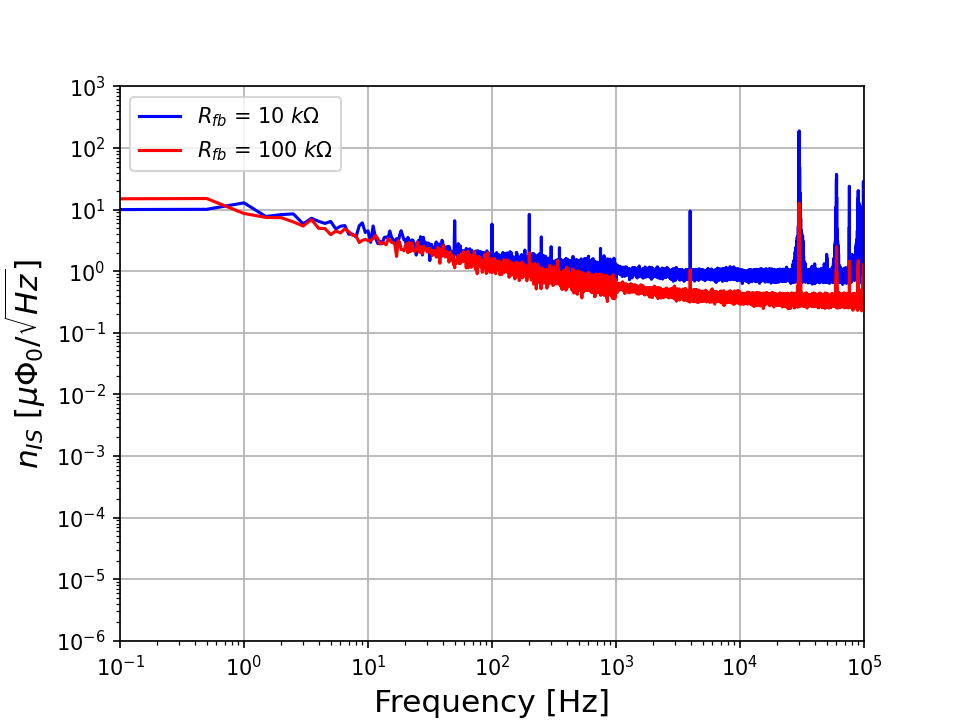}
    \caption{The flux-locked noise of the two-stage dc-SQUID circuit with different feedback resistors at $300$ $mK$.
    \label{fig:fig6_IS_noise}}
\end{figure}

\begin{figure}[htbp]
    \centering
    \includegraphics[width=.7\textwidth]{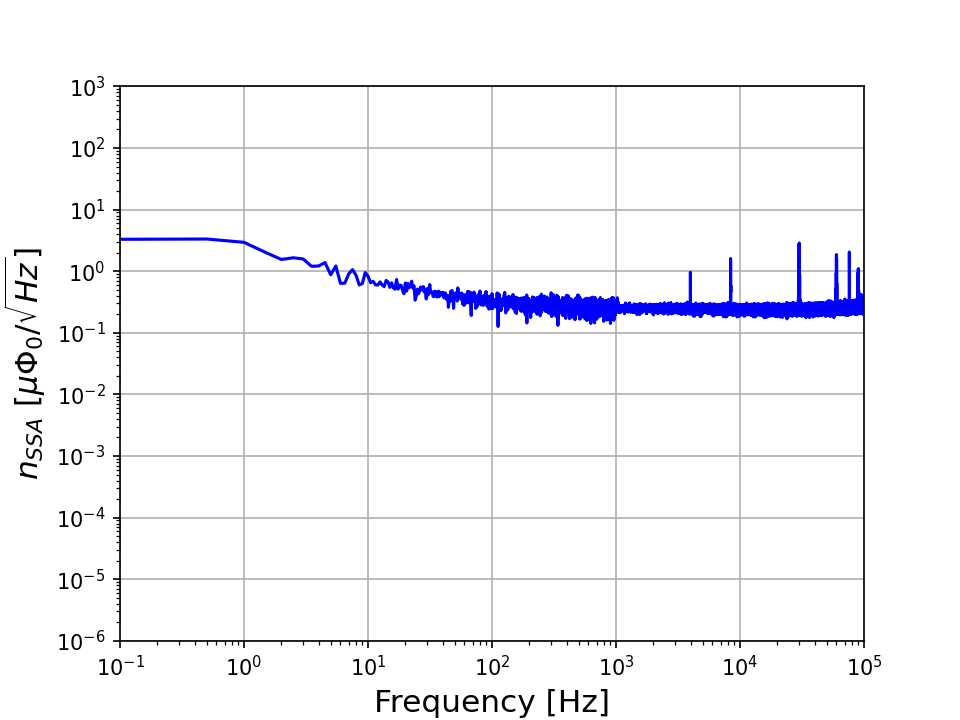}
    \caption{SSA system noise at 300 $mK$ with 100 $k\Omega$ feedback resistor in ADR. 
    \label{fig:fig7_SSA_noise}}
\end{figure}

Figure~\ref{fig:fig6_IS_noise} shows the noise of two-stage dc-SQUID circuit with $10$ $k\Omega$ and $100$ $k\Omega$ feedback resistors separately. 
The flat noise power density is about 1 ${\mu}\Phi_0/\sqrt{Hz}$ at $1$ $kHz$ with $10$ $k\Omega$ feedback resistor.
It is about 0.5 ${\mu}\Phi_0/\sqrt{Hz}$, when the feedback resistor is $100$ $k\Omega$.
This is reasonable.  
According to the linear transfer function of flux-locked loop ($i_{in} = u_{out}/R_{fb}$), the larger the feedback resistance $R_{fb}$, the smaller the influence of noise equivalent current refering to the input coil of input SQUID $i_{in}$ from the voltage of the feedback (or readout) circuit $u_{out}$.
The flat noise power density is about $0.3$ ${\mu}\Phi_0/\sqrt{Hz}$ at 10 $kHz$ after flux-locked and with $100$ $k\Omega$ feedback resistor.
The system noise equivalent current $NEI_{sys}$, which can be calculated by Equation~\ref{eq1}, is about $4$ $pA/\sqrt{Hz}$ at 1 $kHz$ and $2.4$ $pA/\sqrt{Hz}$ at 10 $kHz$, which is significantly lower than the typical noise level of $100$ $pA/\sqrt{Hz}$ for a TES system.
Figure~\ref{fig:fig3_Vphi} shows the dynamic resistance value of the SSA $R_{dyn,SSA}(\Phi)=V_{SSA}(\Phi)/I_{b,SSA}$ at locking point, which is about $50$ $\Omega$. 
$V_{SSA}(\Phi)$ is the voltage swing at different flux input.
$I_{b,SSA}$ is the bias current of SSA.
Thus the total voltage noise of Magnicon electronics $n_{elec}$ is about $0.35$ $nV/\sqrt{Hz}$, which can be calculated by Equation~\ref{eq4}.
And the noise refering to the input coil of input SQUID $NEI_{elec}$ is about $0.17$ $pA/\sqrt{Hz}$, which is negligible.
\begin{equation}
\label{eq4}
    n_{elec} = \sqrt{n_v^2 + (n_iR_{dyn,SSA})^2}.
\end{equation}
We also measured the flux-locked noise of SSA without input SQUID at the same temperature in ADR. 
The intrinsic noise of SSA system is about $0.25$ ${\mu}\Phi_0/\sqrt{Hz}$ at 10 $kHz$, as shown in Figure~\ref{fig:fig7_SSA_noise}.
The $NEI_{SSA}$ is about $1.2$ $pA/\sqrt{Hz}$ refering to input coil of input SQUID based on Equation~\ref{eq2}.
The input SQUID noise equivalent current contribution $NEI_{IS}$ can be calculated to be approximately $2.1$ $pA/\sqrt{Hz}$ at $10$ $kHz$ refering to its input coil.
Therefore, in our two-stage dc-SQUID readout system, the noise contributions from room temperature electronics and even SSA, are negligible.

\section{Conclusion}
\label{sec:conclusion}
We have developed a two-stage dc-SQUID amplifier with exceptionally low noise. 
The system noise equivalent current density is as low as $2.4$ $pA/\sqrt{Hz}$.
The input SQUID exhibits not only an high current sensitive of 8 ${\mu}A/\Phi_0$, but also a substantial flux conversion coefficient $I_{\Phi}{\approx}45$ ${\mu}A/\Phi_0$, which ensures the noise contribution of SSA is merely $1.2$ $pA/\sqrt{Hz}$ refering to the input coil of input SQUID.
The noise contribution of input SQUID is about $2.1$ $pA/\sqrt{Hz}$ refering to its input coil.
The SSA employs a structure with 100 dc-SQUID cells connected in series, achieving a flux conversion coefficient $V_{\Phi}$ of 10 $mV/\Phi_0$.
The room-temperature electronics noise contribution is about $0.17$ $pA/\sqrt{Hz}$, which is negligible.
The two-stage dc-SQUID circuit can satisfy the requirements of TDM readout system, which will be used in AliCPT-40G in the near future. 
Also, it can be applied in detection of X/$\gamma$-rays.


\funding{This work is supported by the National Key Research
and Development Program of China 
(Grant No.2021YFC2203400),
the National Natural Science Foundation of China (Grant No 12173040),
Youth Innovation Promotion Association CAS 2021011,
Scientific Instrument Developing Project of the Chinese Academy of Sciences
(Grant No.YJKYYQ20190065, No.PTYQ2025TD0021),
and the National Key Research and Development Program of China 
(Grant No.2022YFC2204902, No.2022YFC2204900).
}




\bibliographystyle{unsrt}
\bibliography{reference} 

\end{document}